\documentclass[]{aastex63}
\usepackage{listings}
\usepackage{hyperref}

\usepackage{subfigure}
\usepackage{amsmath}
\usepackage{cancel}
\usepackage[version=4]{mhchem}
\usepackage{color, soul}
\setstcolor{red}

\newcommand{\comment}[1]{}

\shorttitle{Accretion flow with nuclear burning}
\shortauthors{Nagarajan and Shigeyama}

\begin{document}
\title{Stationary accretion flow with nuclear burning}
\author{Narenraju Nagarajan}
\author{Toshikazu Shigeyama}
\affiliation{Research Center for the Early Universe (RESCEU), School of Science, The University of Tokyo, 7-3-1 Hongo, Bunkyo-ku, Tokyo 113-0033, Japan}
\date{February 2022}

\begin{abstract}
    We present a series of numerical solutions of spherically symmetric stationary flows with nuclear burning accreted by a neutron star (or black hole). We consider the accretion of matter composed of carbon and oxygen, which mimics the flow after a neutron star is engulfed by a CO star or CO core of a massive star. It is found that there are two types of transonic solutions depending on the accretion rate. The flow with a small accretion rate reaches the center (or the surface of the central object) at supersonic speeds. The other type with a large accretion rate has another sonic point inside the transonic point and the flow truncates at the sonic point. The critical accretion rate dividing these two types is derived as a function of the mass of the central object and the specific enthalpy in the ambient matter. We discuss implications from the solutions for a new mechanism of super-Chandrasekhar type Ia supernovae and type Icn supernovae.
\end{abstract}
\section{Introduction}
Accretion is an important process in close binary systems composed of stars and compact objects. High accretion rates are expected especially in merging events. The tidal disruption of a white dwarf by a neutron star or a black hole was investigated in a series of papers \citep{2012MNRAS.419..827M,2013ApJ...763..108F,2019MNRAS.488..259F} to expect that carbon burning dominates the accretion flow. It is true that the carbon burning is a good candidate to dominate the accretion flow in terms of the energy generation, the ignition temperature, and the associated time scale. The oxygen burning is ignited at too high temperatures or generates too small energy compared with the gravitational energy. {The timescales of both of the nuclear burning is sufficiently short compared with the dynamical timescale of the flow once the temperature exceeds $2\times10^9$ K for carbon burning ($4\times10^9$ K for oxygen burning).} Despite of these features of carbon burning, \citet{2019MNRAS.488..259F} concluded that the carbon burning never dominates the accretion flow originating from the tidal disruption of a white dwarf composed of carbon and oxygen.\\

Here we consider an alternative situation where a neutron star is engulfed by a CO core of a massive star. This may happen in a system similar to the progenitor system of a type Ic supernova (SN Ic) such as SNe 1994I, 1987M\citep{1990A&A...240L...1N, 1994Natur.371..227N}. SN 1994I is considered to be an explosion of a CO star in a close binary system with a neutron star\citep{1994Natur.371..227N}. The hydrogen-rich envelope and helium envelope of the progenitor had been stripped by interaction with the neutron star. The CO star underwent the collapse of the Fe core and exploded as a SN Ic. There is a possibility that the neutron star is engulfed by the CO star before the core collapse if they are more tightly bound and the mass of the CO star is a few times greater than that of the neutron star (a less massive CO star may be disrupted by the tidal force of the neutron star).  The neutron star engulfed by the CO star will accrete carbon-rich matter at extremely high rates. {The accreted matter is expected to have some angular momentum. \citet{1993ApJ...411L..33C} estimated the specific angular momentum of matter accreted by a neutron star in the envelope of a normal star to be $j\sim0.014GM_{\rm N}/c_{\rm s}$, where $G$ denotes the gravitational constant, $M_{\rm N}$ the mass of the neutron star, and $c_{\rm s}$ is the sound speed of the matter. Gas with this angular momentum will eventually move in a circular orbit with a radius of the order of $r\sim10^6$ cm while carbon burning takes place at $r\sim10^8$ cm (see section \ref{results}). Thus the effect of the angular momentum is expected to be unimportant for the nuclear burning.}\\

We investigate how the energy released from carbon burning affects the accretion flow depending on the accretion rate in such situations. In contrast to the approach taken by \citet{2019MNRAS.488..259F} who performed 2D hydrodynamics simulations, we simplify the model as much as possible. We systematically investigate spherically symmetric stationary accretion flows onto neutron stars or black holes with various accretion rates.
On the other hand, we use an equation of state including the ideal gas of ions, radiation, electrons \citep{2000ApJS..126..501T} since \citet{2019MNRAS.488..259F} stressed the importance of an equation of state taking into account the effects of radiation.\\

There is also a possibility that a neutron star engulfed by a He core of a massive star or a He star. The latter situation has been investigated as a possible channel to hydrogen-free luminous supernovae or type Ibn supernovae\citep{2012ApJ...752L...2C}. Furthermore spiral-in of a black hole or a neutron star to a He star is numerically investigated as a possible mechanism of long-duration gamma-ray bursts \citep{1998ApJ...502L...9F,2001ApJ...550..357Z}. Though their simulations took into account relevant nuclear reactions, the results indicated that the energy from nuclear reactions is not crucial in their context. Thus we concentrate on CO cores or CO stars in this paper and leave the accretion of He onto a compact object to our future work.\\

The structure of the paper is as follows. Section 2 describes the governing equations with boundary conditions and introduces our method to solve them. Section 3 presents results from systematic calculations. Section 4 discusses implications from our results and Section 5 concludes the paper.


\section{Formulation of spherically symmetric accretion onto compact objects} \label{SecII}
Suppose that a compact object (a neutron star or a black hole) with a mass $M$ accretes matter at a rate of $\dot M$. We assume that the flow is spherically symmetric and in a stationary state.
Then the governing equations are given as,
\begin{eqnarray}
&\dot M=4\pi r^2 v\rho,& \label{eq:rate}\\
&\rho v\frac{dv}{dr}+\frac{dp}{dr}+\frac{GM\rho}{r^2}=0,&\\
&v\left(\frac{d\epsilon}{dr}-\frac{p}{\rho^2}\frac{d\rho}{dr}\right)=\varepsilon_{\rm nuc}(\rho,T).&\label{eq:energy}
\end{eqnarray}
Here $r$ denotes the radial coordinates with the origin located at the position of the compact object (a neutron star or a black hole), $v$ the radial velocity, $\rho$ the mass density, $p$ the pressure, $\epsilon$ the specific thermal energy density,  $\varepsilon_{\rm nuc}(\rho,T)$  the energy generation rate due to nuclear reactions as a function of the density and the temperature $T$,  and $G$ is the gravitational constant. { We ignore the cooling term due to neutrino emission. We will discuss the effects of the neutrino cooling referring to our results in section \ref{types}.} Note that the accretion rate $\dot M$ and the velocity $v$ have negative values.\\

After some manipulations, we convert these three equations into the following form convenient for numerical integration.

\begin{eqnarray}
&\frac{d\rho}{dr}=\frac{\rho^2 \left((\partial\epsilon/\partial p)_\rho GM \rho v-2 (\partial\epsilon/\partial p)_\rho r \rho v^3+r^2 \varepsilon_{\rm nuc}\right)}{r^2 v \left((\partial\epsilon/\partial p)_\rho \rho^2 v^2+(\partial\epsilon/\partial\rho)_p \rho^2-p\right)},&\label{drhodr} \\ 
&\frac{dv}{dr}= \frac{(\partial\epsilon/\partial p)_\rho GM \rho^2 v+2 (\partial\epsilon/\partial\rho)_p r \rho^2 v+r^2 \varepsilon_{\rm nuc} \rho-2 r p v}{r^2 \left(-(\partial\epsilon/\partial p)_\rho \rho^2 v^2-(\partial\epsilon/\partial\rho)_p \rho^2+p\right)},& \label{dvel}\\
&\frac{dp}{dr}= \frac{(\partial\epsilon/\partial\rho)_p GM \rho^3-2 (\partial\epsilon/\partial\rho)_p r \rho^3 v^2-r^2 \varepsilon_{\rm nuc} \rho^2 v-GMp\rho+2 r p \rho v^2}{r^2 \left(-(\partial\epsilon/\partial p)_\rho \rho^2 v^2-(\partial\epsilon/\partial\rho)_p \rho^2+p\right)}.& \label{dpre}
\end{eqnarray}

We numerically solve these equations using the Runge-Kutta method with the equation of state given by \citet{2000ApJS..126..501T} under the boundary conditions described in the next section.
\subsection{Transonic point} \label{sonic_point}
The matter surrounding the central object is originally in hydro-static equilibrium under the gravity of a CO star. Thus the accreted matter moves at subsonic speeds far from the central compact object and gradually accelerates toward the central object due to its gravity. 
At the transonic point ($r=r_\mathrm{s}$), the denominator of the above equations vanishes. Since the accretion flow under consideration smoothly passes through the transonic point and accelerates to supersonic speeds to reach the central object, we require the numerator also vanishes simultaneously. Thus the following equations should hold at $r=r_\mathrm{s}$.
\begin{eqnarray}
&GM-2v^2r=0,&\label{kepler}\\
&(\partial\epsilon/\partial p)_\rho GM\rho^2+2(\partial\epsilon/\partial\rho)_p r \rho^2-2 r p=0.& \label{den}
\end{eqnarray}
From the former equation (\ref{kepler}), we obtain
\begin{equation}\label{vel}
    v=-\sqrt{\frac{GM}{2r}}.
\end{equation}
The total energy flux can be derived from the three governing equations (\ref{eq:rate})--(\ref{eq:energy}) as 
\begin{equation}
    4\pi r^2v\left(\frac{\rho v^2}{2}+\rho\epsilon+p-\frac{GM\rho}{r}\right),
\end{equation}
which should be constant unless nuclear energy is generated. Thus we can equate this quantity with the sum of the value at a distant point $r=r_\infty$ where the gravity and the kinetic energy are negligible. The result can be expressed as 
\begin{equation}\label{fene}
    4\pi r^2v\left(\frac{\rho v^2}{2}+\rho\epsilon+p-\frac{GM\rho}{r}\right)=\dot M h_\infty+\dot E_\mathrm{nuc},
\end{equation}
where $h_\infty$ is the specific enthalpy at $r=r_\infty$ in the ambient matter and $\dot E_\mathrm{nuc}=4\pi\int r^2\rho\varepsilon_\mathrm{nuc}dr$.
For given $\dot M$ and $h_\infty$ we obtain the values of $r_\mathrm{s}$, $v$, $\rho$, and $p$ at the transonic point from equations  (\ref{eq:rate}), (\ref{den}), (\ref{vel}), (\ref{fene}), and the equation of state if carbon is not ignited outside the transonic point, i.e., $\dot E_\mathrm{nuc}=0$. We assume that this is the case when we derive the quantities at the transonic point. After carbon is ignited inside the transonic point, another sonic point due to a finite $\dot E_\mathrm{nuc}$ may appear depending on the accretion rate. \\

We need to start numerical integration from a point slightly off the transonic point. To do this we evaluate the derivatives of $\rho$, $v$, and $p$ at the transonic point. Substitutions of the Taylor expansions of these variables around the transonic point into equations (\ref{drhodr})--(\ref{dpre}) yield equations for the derivatives. Applications of Rolle's theorem to equations  (\ref{drhodr})-- (\ref{dpre}) yield the same results.  The resultant equations for the derivatives of $\rho$ and $p$ ($\rho^\prime$, $p^\prime$) with respect to $r$ are as follows:
\begin{eqnarray}
&\rho^\prime=& \nonumber \\
&\frac{2 GM \left(\frac{\partial\epsilon}{\partial_p}\right)_\rho \rho^3 \left(2 r \rho'+3 \rho\right)}{r \left(r \rho p' \left(\rho^2\left(\left(\frac{\partial^2\epsilon}{\partial p^2}\right)_\rho GM+2 \frac{\partial^2\epsilon}{\partial\rho\partial p} r\right)-2 r\right)+r \rho' \left(\rho^2\left(\frac{\partial^2\epsilon}{\partial\rho\partial p} GM \rho+2 r\left(\left(\frac{\partial^2\epsilon}{\partial \rho^2}\right)_p \rho+\left(\frac{\partial\epsilon}{\partial_\rho}\right)_p\right)-GM \left(\frac{\partial\epsilon}{\partial p}\right)_\rho\right)+2 r p\right)-4 GM \left(\frac{\partial\epsilon}{\partial_p}\right)_\rho \rho^3\right)}& \label{drho},\\
&p^\prime=& \nonumber \\
&\frac{2 GM \left(2 r\rho'+3 \rho\right) \left(p-\left(\frac{\partial\epsilon}{\partial\rho}\right)_p\rho^2\right) }{r \left(r \left(p' \left(\rho^2 \left(\left(\frac{\partial^2\epsilon}{\partial p^2}\right)_\rho GM+2 \left(\frac{\partial^2\epsilon}{\partial p\partial\rho}\right) r\right)-2 r\right)+\rho^2 \rho'\left(\left(\frac{\partial^2\epsilon}{\partial p\partial\rho}\right) GM+2 \left(\frac{\partial^2\epsilon}{\partial\rho^2}\right)_p r\right)+4 \left(\frac{\partial\epsilon}{\partial\rho}\right)_p \rho \left(r \rho'+\rho\right)-4 p\right)-2 GM \left(\frac{\partial\epsilon}{\partial p}\right)_\rho \rho^2\right)}.& \label{dp}
\end{eqnarray}
Here the derivative of $v$ ($v^\prime$) has been already eliminated using 
\begin{equation}\label{dv}
    v^\prime/v+\rho^\prime/\rho+2/r=0.
\end{equation}
These are quadratic equations for $\rho^\prime$ ($p^\prime$). Note here that $\rho, \, p, \, v$, and $r$ are those that have been obtained at the transonic point $r=r_\mathrm{s}$. We should take the solutions that give a supersonic (subsonic) flow inside (outside) the sonic point. 

\subsection{Nuclear Reactions} \label{nuclear_reaction}
Hydrogen burning in the central region of a star yields helium, and the subsequent helium burning gives rise to the formation of the core composed of $\ce{^12_6C}$ and $\ce{^16_8O}$. We suppose that a neutron star engulfed by this core accretes the ambient matter and that the carbon starts to undergo fusion. We include the following formulae for nuclear burning rates of carbon and oxygen as additional governing equations into the numerical integration procedure alongside equations (\ref{drhodr}), (\ref{dvel}), and (\ref{dpre}).\\

We utilise the nuclear burning rates used in \citet{2019MNRAS.488..259F}, which takes into account the following reaction as the primary carbon-burning process since it is the most energetic among the $\alpha$ reactions involving carbon, oxygen and helium.
\begin{equation} \label{carbon_fusion}
    \ce{^12_6C + ^12_6C -> ^24_12Mg + \gamma}+ 13.933\,\rm MeV
\end{equation}

The rate of change of the mass fraction $X_\mathrm{C}$ of $\ce{^12_6C}$ is given by,
\begin{equation} \label{xdot_carbon}
    \dot X_{\rm C} =v\frac{dX_{\rm C}}{dr}= -\frac{m_\mathrm{C}}{Q_{12}}\dot Q_{\rm nuc, 12}
\end{equation}
where $m_{\rm C}$ is the mass of the carbon nucleus, $Q_{12}=13.933$ MeV is the energy released in the nuclear reaction (\ref{carbon_fusion}), and the specific energy generation rate $\dot Q_{\rm nuc, 12}$ is approximated as,
\begin{equation}
    \dot Q_{\rm nuc, 12} = 3.96\times10^{43}\,\rho\,X_C^2\,\frac{T_{A9}^{5/6}}{T_{9}^{3/2}}\times \exp{\left(-\frac{84.165}{T_{\rm A9}^{1/3}}-2.12\times10^{-3}\,T_9^3\right)}\,{\rm erg\,g^{-1}\,s^{-1}}
\end{equation}
where $\rho$ is the density in g\,cm$^{-3}$. The variables $T_9$ and $T_{\rm A9}$ are defined as,
\begin{eqnarray}
    T_9 &=& \frac{T}{10^9\,\rm K},\\
    T_{\rm A9} &=& \frac{T_9}{1+0.0396\,T_9}
\end{eqnarray}
where $T$ is the temperature in units of K.\\

Based on recent experiments, we also take into account the increase in $\ce{^12_6C}\,+\,\ce{^12_6C}$ fusion rate due to resonances \citep{TUMINO2018} at the suggested temperature ranges.  The analytical function used to describe the updated nuclear reaction rate is given by,
\begin{equation}
    N_A\left<\sigma v\right> = \Sigma_{i=1}^{3} \exp{\left[ a_{i1}+a_{i2}T_9^{-1}+a_{i3}T_9^{-1/3}+a_{i4}T_9^{1/3}+a_{i5}T_9+a_{i6}T_9^{5/3}+a_{i7}\ln{(T_9)}\right]}
\end{equation}
where the coefficient $a_{ij}$ is taken from Table 1 in \citet{TUMINO2018}. However, this phenomenon should not affect the results for temperatures above $\approx2\times10^9$ K. We confirmed that essentially the same solutions are obtained using the two different reaction rates.

We follow \citet{CAUGHLAN1988283}, for the nuclear reaction rates of oxygen. This is governed by the fusion reaction,
\begin{equation} \label{oxygen_fusion}
    \ce{^16_8O + ^16_8O -> ^32_16S + \gamma}+ 16.000\,\text{MeV}
\end{equation}
The rate of change of the mass fraction $X_\mathrm{O}$ of $\ce{^16_8O}$ is addressed similar to equation (\ref{xdot_carbon}) as,
\begin{equation} \label{xdot_oxygen}
    \dot X_{\rm O} =v\frac{dX_{\rm O}}{dr}= -\frac{m_{\rm O}}{Q_{16}}\dot Q_{\rm nuc, 16}
\end{equation}
where $m_{\rm O}$ is the mass of the oxygen nucleus, $Q_{16}=16.000$ MeV is the energy released in the nuclear reaction (\ref{oxygen_fusion}), and the specific energy generation rate $\dot Q_{\rm nuc, 16}$ is defined as,{
\begin{equation}
\dot Q_{\rm nuc, 16} = \frac{2.14\times10^{53}\rho X_\mathrm{O}^2}{T_{93}^2} \times \exp{\left(\frac{-135.93}{T_{93}}-0.629\,T_{93}^2-0.445\,T_{93}^4+0.0103\,T_9^2\right)}\,\mathrm{erg\,g}^{-1}\,\mathrm{s}^{-1},
\end{equation}
}
where $T_{93}$ is defined as $T_{93} = T_9^{1/3}$.
When the governing equations for the flow are integrated we substitute these energy generation rates $\dot Q_{\rm nuc, 12}$ and $\dot Q_{\rm nuc, 16}$ into $\varepsilon_{\rm nuc}(\rho,T)$ as
\begin{equation}
    \varepsilon_{\rm nuc}(\rho,T)=\dot Q_{\rm nuc, 12}+\dot Q_{\rm nuc, 16}.
\end{equation}

\subsection{Numerical integration} \label{NumericalIntegration}
For a given set of $h_\infty$, $\dot M$, and the mass fractions of carbon and oxygen in the ambient matter, we start the integration of differential equations (\ref{drhodr})--(\ref{dpre}) together with equations (\ref{xdot_carbon}) and (\ref{xdot_oxygen}) from a point with a radius $r_\mathrm{o}=r_\mathrm{s}+\delta r$ ($\delta r>0$) to a large $r(>>r_\mathrm{s})$ where the gravitational energy can be neglected compared with the internal energy of the gas in equation (\ref{fene}). Here $r_\mathrm{s}$ denotes the radius of the transonic point given by solving equations (\ref{kepler})--(\ref{den}).  The initial values of $\rho$, $p$, and $v$ are given by
\begin{equation}
\rho(r_\mathrm{o})=\rho(r_\mathrm{s})+\rho^\prime(r_\mathrm{s})\delta r,\, p(r_\mathrm{o})=p(r_\mathrm{s})+p^\prime(r_\mathrm{s})\delta r,\, 
v(r_\mathrm{o})=v(r_\mathrm{s})+v^\prime(r_\mathrm{s})\delta r.
\end{equation}
The flow in this outer part must be subsonic to meet the present context. We set $X_\mathrm{C}=0.3$ and $X_\mathrm{O}=0.7$ in the ambient matter as fiducial values in the CO core of a massive star throughout the paper.  Here we have assumed that no nuclear reaction takes place in this region(that is, $\dot E_\mathrm{nuc}=0$ in Eq. (\ref{fene})), which is confirmed to a good accuracy as far as the temperature in the ambient matter is below $\sim4\times10^8$ K. Thus the integrations of equations (\ref{xdot_carbon}) and (\ref{xdot_oxygen}) are started with these values to both sides of the transonic point.  \\

Subsequently, we integrate differential equations (\ref{drhodr})--(\ref{dvel}) from a point with $r_\mathrm{i}=r_\mathrm{s}-\delta r$ toward the center. The initial values are given by
\begin{equation}
\rho(r_\mathrm{i})=\rho(r_\mathrm{s})-\rho^\prime(r_\mathrm{s})\delta r,\, p(r_\mathrm{i})=p(r_\mathrm{s})-p^\prime(r_\mathrm{s})\delta r,\, 
v(r_\mathrm{i})=v(r_\mathrm{s})-v^\prime(r_\mathrm{s})\delta r.
\end{equation}


\begin{figure}[htbp]
 \centering
 \includegraphics[width=1.0\linewidth]{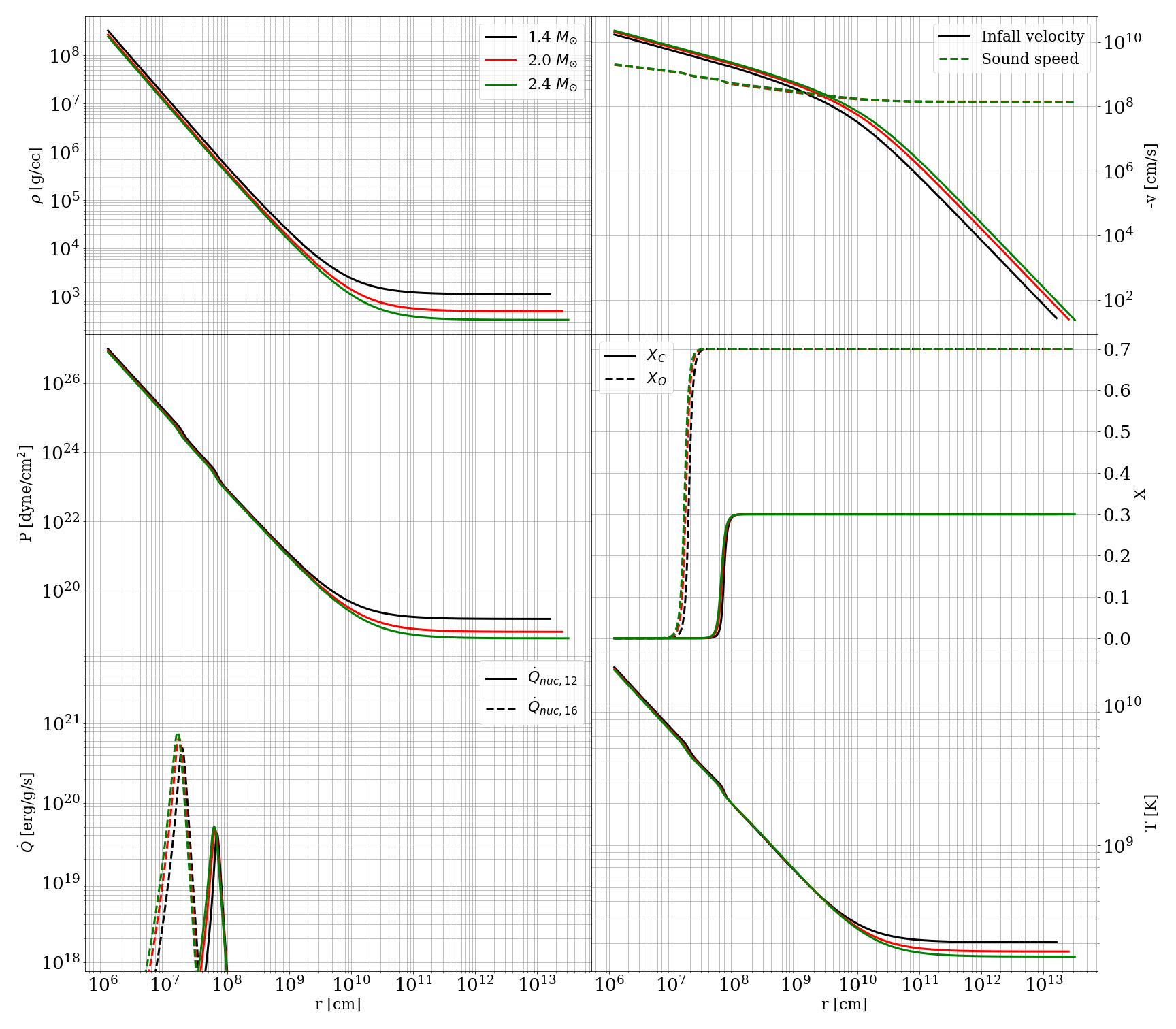}
\caption{Structures of accretion flows as functions of radius for 3 different masses of compact objects as the labels in the top left panel indicate. The accretion rate is $\dot M=-1.0\times 10^{32}$ g s$^{-1}$. The other parameters in the ambient matter are $\log h_\infty=16.6025$, $X_\mathrm{C}=0.3$, and $X_\mathrm{O}=0.7$. In the left panels, the density, the pressure, and the energy generation rates are shown from the top to the bottom. In the right panels, the top panel shows the sound speed (dashed line) and the infall velocity (solid lines), the middle panel the mass fractions of C (solid line) and O (dashed line), and the bottom panel shows the temperature.}
\label{fig:smallmdot}
 \end{figure}
\section{Results} \label{results}
\subsection{Two types of solutions}\label{types}
In general, the accretion flow that does reach the surface of the central object is monotonically accelerated toward the center (Fig. \ref{fig:smallmdot}) though the flow shows some influence from the nuclear reactions. We observe that each of the nuclear energy generation rate has a peak in the supersonic inner region. An interesting feature seen from this figure is that the maximum value of $\varepsilon_\mathrm{nuc}$  decreases with increasing mass for carbon burning but the opposite is true for the oxygen burning. The energy generation rate of each burning has a peak at the same temperature while the densities at the peaks depend on the mass of the central object. The location of the peak of the energy generation rate for carbon burning is not sensitive to the mass of the central object. Thus the energy generation rate for carbon burning has a lower peak for a higher mass of the central object for which the density becomes shallower at the same radii. On the other hand, the location of the peak of oxygen burning is significantly shifted depending on the mass of the central object due to higher infall velocity. The peak is attained at a higher density for a higher mass of the central object, thus leading to a higher peak.  

In reality, the continuous increase in temperature toward the center implies further nuclear reactions {and neutrino emission}.  We have ignored the nuclear reactions at higher temperatures because the energy generation due to nuclear reactions within a region with $r<10^8$ cm is too small compared with the gravitational energy to affect the flow. {We estimated the cooling rate due to neutrino emission using rates available from the literature \citep[e.g., ][]{1996ApJS..102..411I} around the region where carbon burning takes place. We found that the cooling rate is of the order of $10^{16}$ erg g$^{-1}$ s$^{-1}$ at maximum in the upstream of the burning region and that the cooling time scale is at least a factor of 10 longer than the advection timescale. In the burning region, the heating term dominates the cooling term by more than two orders of magnitudes. Thus it is expected that the neutrino cooling affects the flow to some extent. If more accurate quantities are necessary, one needs to include not only this cooling term but also more extensive nuclear reaction networks. These are issues for the future study.}

Moreover, it is expected that an accretion shock is generated when the supersonic flow reaches the surface of the central object. {Actually, \citet{2005ApJ...623.1000Y} estimated the radius of the accretion shock for the spherically symmetric flows in the context of Fe core collapse of massive stars. The results suggest that the accretion shock for stable solutions extends up to $\sim3\times10^7$ cm at most. This means the existence of the accretion shock does not affect our results on the critical accretion rates, which are determined by carbon burning taking place at $r\sim10^8$ cm. Thus the critical accretion rates are not affected by neutrino emission for the same reason.}

By performing a series of calculations with different $\dot M$'s, we find that there exist certain solutions in which the flow meets another sonic point after undergoing carbon burning inside the transonic point (Fig. \ref{fig:largemdot}).  These solutions cannot be integrated further and thus do not reach the surface clearly due to the carbon burning. For ambient matter with the same specific enthalpy, this type of solutions have larger accretion rates (in terms of $|\dot M|$). {These solutions suggest that the oxygen burning does not dramatically change the characteristics of the accretion flow because the ignition temperature is so high that the energy from the oxygen burning is available only in a deep gravitational potential where the gravitational energy dominates over the nuclear energy}. 

{To illustrate the effects of nuclear burning, we compare two of these solutions with and without nuclear burning in Figure \ref{fig:effects_of_nb}. One solution with a large accretion rate does not reach the center due to nuclear burning and the other solution with a smaller accretion rate reaches the surface of the central object. This figure shows that energies generated from both of carbon burning and oxygen burning enhance the pressure of the flow in solutions for both accretion rates. If there is no nuclear burning, a sudden increase in pressure toward the downstream is not observed. The right panel clearly shows that the nuclear energy generation in a solution with a large accretion rate increases the sound speed and truncates the supersonic flow at another sonic point (solid line in the right panels).}

These qualitative features are shared by solutions with different masses of the central object ranging from 1.4 $M_\odot$ to 2.4 $M_\odot$ (Fig. 1), though we do not investigate more massive ones.

If we compare solutions with different accretion rates (see Fig. \ref{fig:comparison_w_different_mdot}), the place where the specific energy generation rate $\varepsilon_\mathrm{nuc}$ attains the maximum value shifts to the outer radius with increasing mass. This is due to higher temperatures attained for higher accretion rates with the same $h_\infty$. Flows with higher accretion rates are affected more significantly by nuclear burning though these solutions have rates smaller than the critical accretion rate. 
 
\subsection{Critical accretion rates}
We have systematically obtained solutions for different specific enthalpies $h_\infty$'s and accretion rates $\dot M$'s to search the critical accretion rate that separates these two different types of solutions as a function of $h_\infty$ in the following manner. We search the parameter space within the following values or range of values,

\begin{enumerate}
	\item Mass of the central object $M=\{1.4,\,2.0,\,2.4\}\,M\textsubscript{\(\odot\)}$.
	\item Specific enthalpy in the ambient matter, $\log h_\infty \ {({\rm erg\ g^{-1}})}= \left[16.1,\,17.1\right]$.
	\item Accretion rate, $\dot M = \left[-10^{31},\,-10^{35}\right]\,$ g s$^{-1}$.
\end{enumerate}
Here we have chosen the range of specific enthalpy to cover the values taken by the matter in the CO cores of massive stars.  
For the sake of reproducible results, we also present here the resolution parameters of the numerical integration module. 
\begin{enumerate}
	\item Spatial resolution for numerical integration attains $\approx 3\times 10^3$ cm near the center. This value was chosen by performing a systematic resolution study such that all relevant physics can be represented with negligible error.
	\item Resolution of $\dot M$ chosen to search for the critical mass accretion rate $\approx 2.0\times 10^{31}$ g s$^{-1}$.
	\item Resolution of $h_{\infty}$ for a given mass of neutron star $\delta h_\infty/h_\infty\approx 10^{-2}$
\end{enumerate}

Searching through the above parameter space shows that there indeed exists a hard boundary above which the flow never reaches the central object. Figure \ref{fig:criticalmdot} depicts this so-called critical accretion rate for different masses of the central object. 
Given the governing equations (\ref{drhodr}), (\ref{dvel}) and (\ref{dpre}) alongside the rate of change of mass fractions (\ref{xdot_carbon}) and (\ref{xdot_oxygen}), we search through the \textit{parameter space} to identify solutions that \textit{reach the central object}. In order to test the hypothesis regarding $h_\infty$ in Sec.\ref{NumericalIntegration}, we set values of the density and temperature in the ambient matter that result in a log uniform distribution in $h_\infty$. Then, for a particular mass of the central object, we vary the accretion rate over a large enough range to search for the existence of a critical value. Consolidating the procedure given in subsections \ref{sonic_point}, \ref{NumericalIntegration} and \ref{nuclear_reaction}: (i) we compute physical quantities at the transonic point given an input accretion rate, enthalpy, and compositions, (ii) perturb these parameters by a small amount using equations (\ref{drho}), (\ref{dp}), and  (\ref{dv}), and (iii) integrate from the perturbed transonic point towards the surface. We then weed out any solutions that do not reach the surface.
Thus we construct Figure \ref{fig:criticalmdot} that shows the critical accretion rates for three different masses of the central object as functions of $h_\infty$.
\begin{figure}[htbp]
 \centering
 \includegraphics[width=1.0\linewidth]{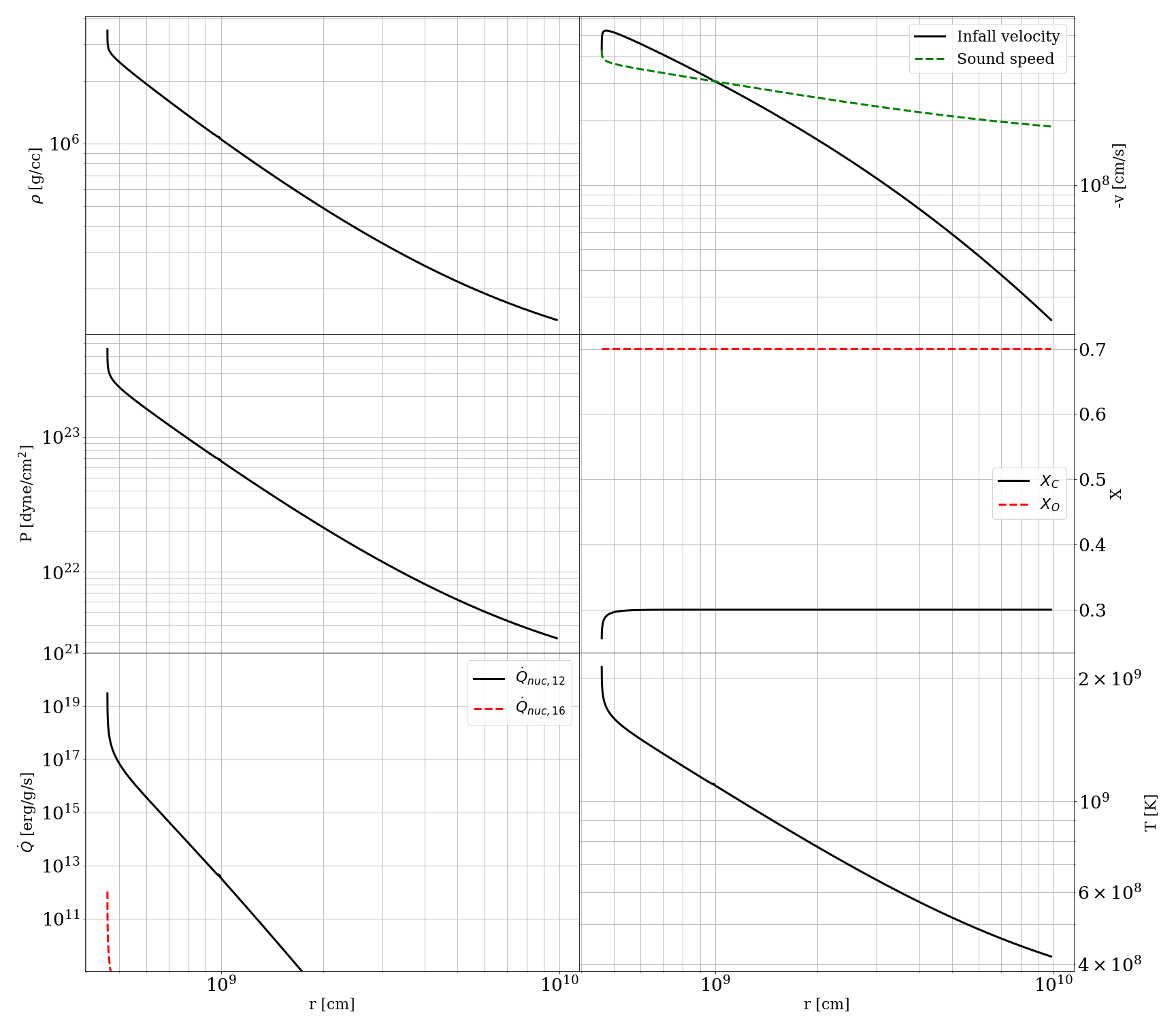}
\caption{The same as Figure \ref{fig:smallmdot} but for $M=1.4\ M_\odot$ and $\dot M=-4.0\times 10^{32}$ g/s. The figure focuses on the inner part as compared with Figure \ref{fig:smallmdot} because the flow truncates at another sonic point inside the transonic point due to infinite derivatives of physical quantities with respect to radius.}
 \label{fig:largemdot}
 \end{figure}

\begin{figure}[htbp]
 \centering
 \includegraphics[width=1.0\linewidth]{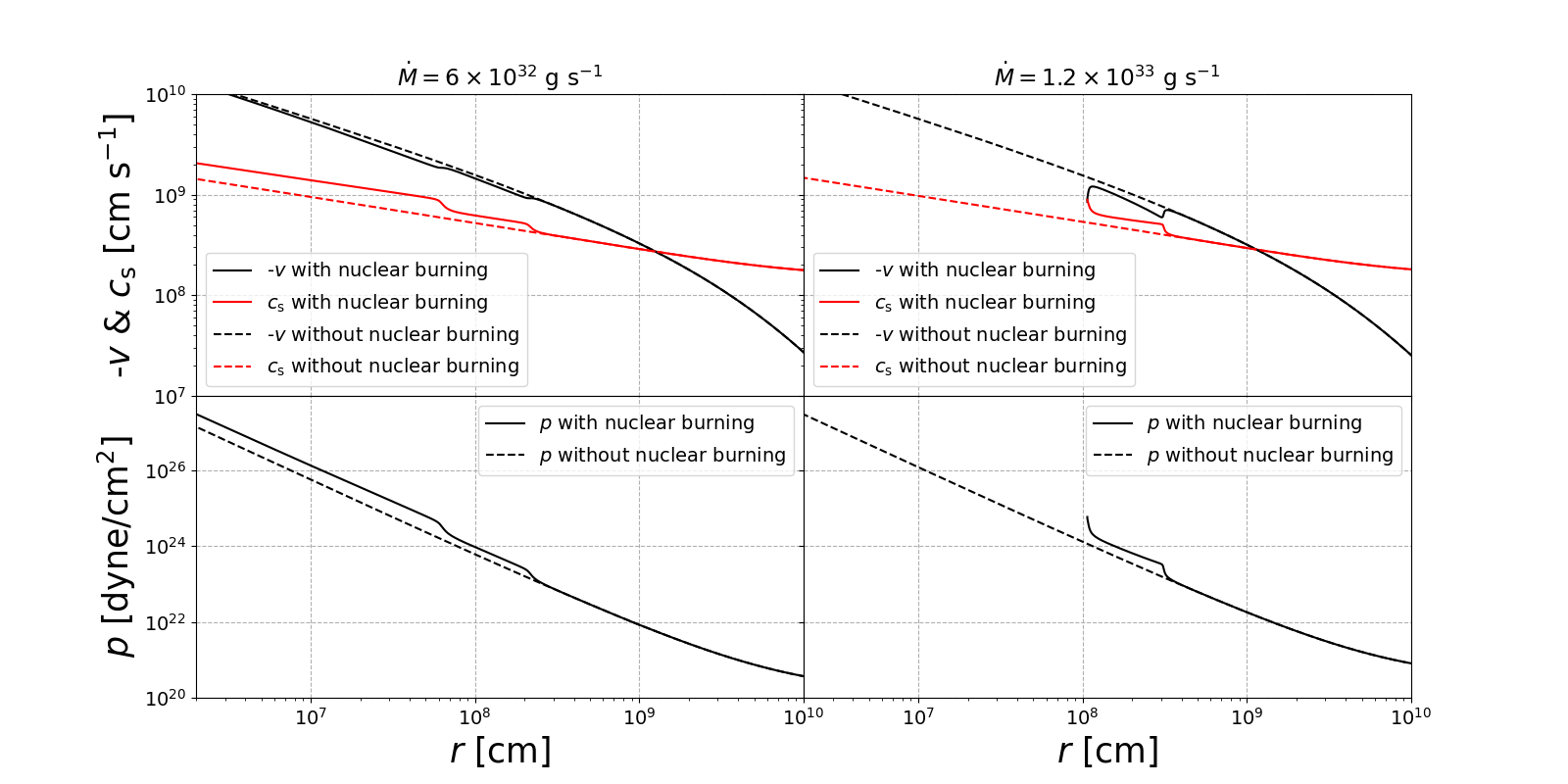}
\caption{{Structures of accretion flows as functions of radius with (solid lines) and without (dashed lines) nuclear burning. The left and right panels show models with accretion rates of $\dot M=-6.0\times 10^{32}$ g s$^{-1}$ and $\dot M=-1.2\times 10^{33}$ g s$^{-1}$, respectively. The other parameters in the ambient matter are $\log h_\infty\ [{\rm erg\ g^{-1}}]=16.6025$, $X_\mathrm{C}=0.3$, and $X_\mathrm{O}=0.7$. The top panels show the infall velocity and the sound speed and the bottom panel the pressure.}}
\label{fig:effects_of_nb}
 \end{figure}
 
We find that the temperature (in the range of $10^8<T_\infty(\rm K)<4\times10^8$) of the ambient matter does not affect the values of the critical accretion rate as a function of $h_\infty$. We note that the critical accretion rate bents at $h_{\infty}\sim4\times10^{16}$ erg g$^{-1}$. This point indicates the transition from the radiation dominated matter to the gas dominated matter with increasing $h_\infty$, which increases the adiabatic index and thus reduces the necessary accretion rate to ignite carbon burning. This figure also depicts ranges of specific enthalpies and accretion rates onto the compact object with a mass of $M=1.4\ M_\odot$ in the companion CO cores of different masses. The structure of CO cores are calculated using the stellar-evolution code MESA (r15140) \citep{Paxton11, Paxton13, Paxton15, Paxton18, Paxton19}. The accretion rates are estimated from the formula $4\pi(GM)^2\rho c_\mathrm{S}^{-3}$ at each cell of the models, where $c_\mathrm{S}$ denotes the sound speed. In reality, the internal structure should be changed from that of a single star after engulfing a compact object.  In addition, more massive stars have less carbon and more oxygen with $X_{\rm C}$ smaller than 0.3 assumed in the derivation of the critical accretion rate. Furthermore the scale of the CO cores is at most $\sim 10^{11}$ cm while our models indicate that the physical quantities become nearly uniform at $>10^{12}$ cm. Thus our models might be too simple to be directly applied to real situations. On the other hand, the transonic points and the burning points are located at $<10^9$ cm, much smaller than the scale of the star.  It is important to note here that there exists a possibility for a compact object engulfed by a lower mass CO companion accretes matter at a rate above the critical accretion rates because the critical accretion rates are comparable to the expected accretion rates. In these systems, the accretion should never reach the central object and might result in a dynamical phenomenon driven by a detonation wave. To explore the results, we need to perform hydrodynamic calculations including nuclear energy generations, which is beyond the scope of this paper.\\

The order of magnitude of the critical accretion rate can be estimated from the following considerations.
The critical accretion rate is defined as the accretion rate above which the accretion flow is affected by the energy generation from nuclear burning and meets another sonic point where the flow truncates. {This situation is realized where the energy generated from the nuclear burning within accretion (or advection) timescale becomes comparable to the gravitational energy. A similar argument was made by \citet{2009ApJ...694..664M} for neutrino heating in accreting neutron stars.} That is,
\begin{equation}
	\varepsilon_{\rm nuc}(\rho,T)\,\delta t \approx \frac{GM}{r_{\rm ig}},
\end{equation}
holds. Here, $\delta t$ denotes the accretion timescale that can be estimated as a fraction of the dynamical timescale by,
\begin{equation}
	\delta t = \frac{-f\times r_{\rm ig}}{v} = f\times \sqrt{\frac{r_{\rm ig}^3}{2GM}},
\end{equation}
and $r_{\rm ig}$ denotes the radius at which the energy generation has a maximum value. The factor $f$ is a constant of the order of unity. 

 \begin{figure}[htbp]
    \centering
    \includegraphics[width=1.0\linewidth]{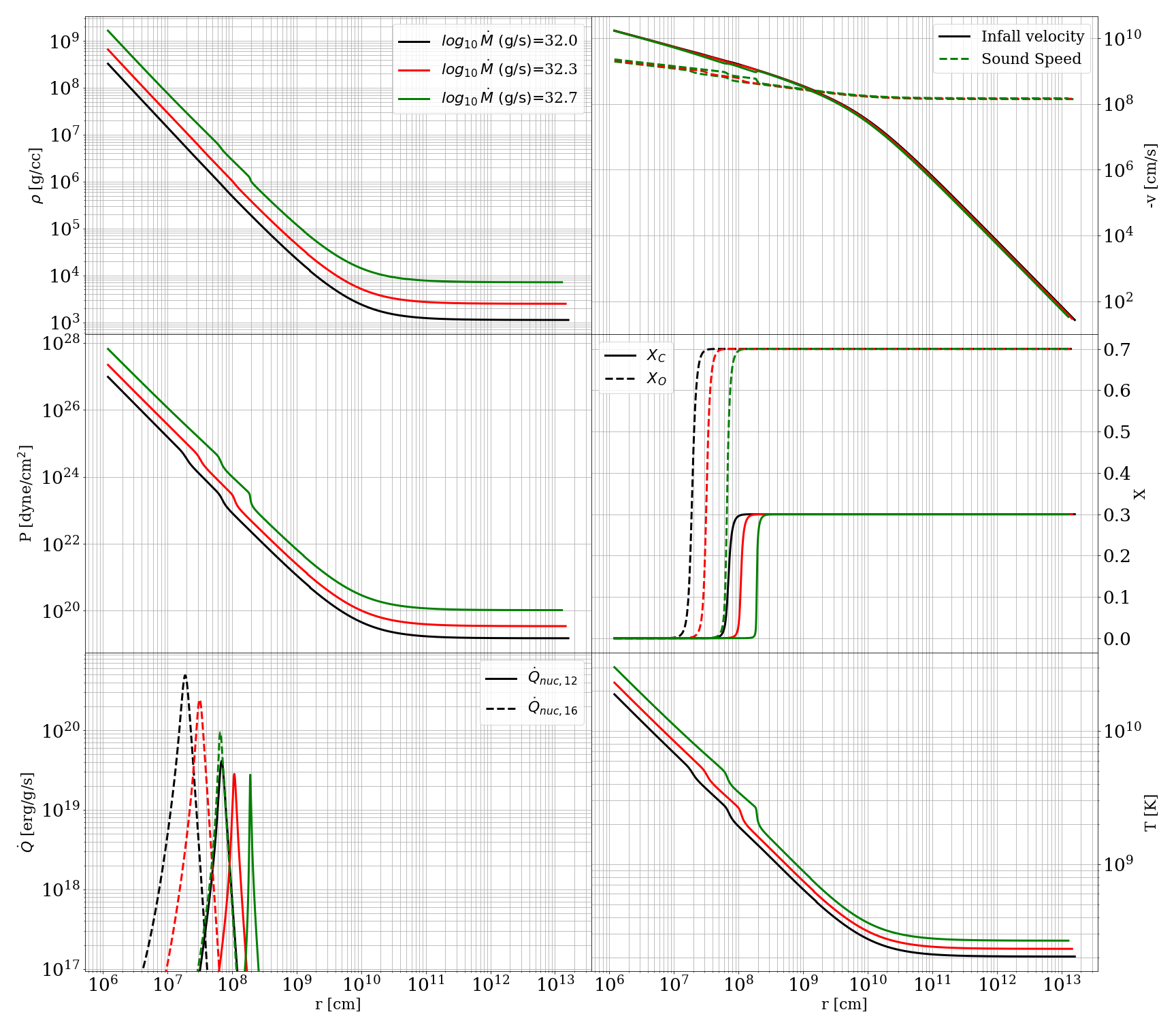}  
    \caption{The same as Figure \protect\ref{fig:smallmdot} but comparing solutions with different accretion rates indicated in the labels in the top left panel. These solutions have the same mass ($M=1.4\ M_\odot$) of the central object and the same specific enthalpy ($\log h_\infty\ {({\rm erg\ g^{-1}})}=16.6025 $) in the ambient matter.}
    \label{fig:comparison_w_different_mdot}
\end{figure}
The density can be expressed as a function of $r_{\rm ig}$ using the critical accretion rate $\dot M_\mathrm{c}$ as,{
\begin{equation}
	\rho \approx \frac{-\dot M_{\rm c}}{4\pi r_{\rm ig}^2\alpha}\sqrt{\frac{r_{\rm ig}}{2GM}}.
\end{equation}
Here $\alpha$ is a constant that accounts for the finite enthalpy in the ambient medium and usually ranges from 1/2 to 1 \citep[e.g.,][]{1990RvMP...62..801B}}. Eliminating $\delta t$ and $\rho$ from the above three equations, $\dot M_\mathrm{c}$ can be expressed as,
{
\begin{equation}
	-\dot M_\mathrm{c} \approx \frac{8\pi\alpha \left(GM\right)^2{\rho}}{fr_{\rm ig}\varepsilon_{\rm nuc}(\rho,T)}\approx1.8\times10^{32}\,{\rm g\, s}^{-1}\frac{\alpha}{f}\left(\frac{M}{1.4\, M_\odot}\right)^2\left(\frac{\dot Q_{\rm nuc,\ 12}/\rho}{10^{13}\, {\rm erg\, g^{-1}\,
s^{-1}}}\right)^{-1}\left(\frac{r_{\rm ig}}{4\times10^8\, {\rm cm}}\right)^{-1}
\end{equation}
}
Here note that the location $r_{\rm ig}$ as well as the temperature distribution is insensitive to $M$ from actual solutions shown in Figures \ref{fig:smallmdot} and \ref{fig:largemdot} and that $\dot Q_{\rm nuc,\ 12}/\rho$ is independent of $\rho$. Thus the critical accretion rate is proportional to the square of $M$, which reproduces the feature of the critical accretion rates shown in Figure \ref{fig:criticalmdot}. Though the order of magnitude of the critical accretion rate and its dependence on the mass of the central object is reproduced with $f\sim0.1$, the actually calculated values depend on the specific enthalpy in the ambient matter, which cannot be reproduced by this approach.\\

\begin{figure}[htbp]
 \centering
 \includegraphics[width=0.7\linewidth]{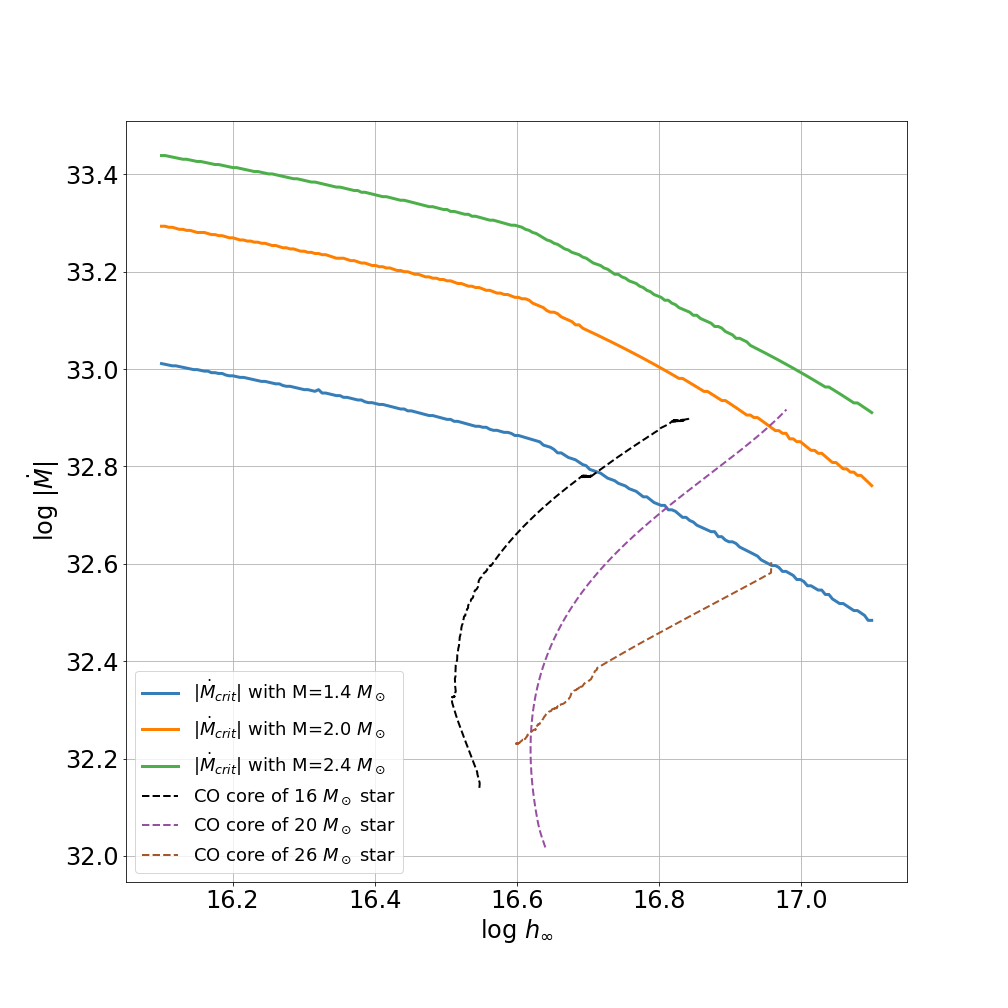}
\caption{The logarithm of the absolute value of critical accretion rates in units of $M_\odot$ s$^{-1}$ as functions of the specific enthalpy {(erg\ g$^{-1}$)} in the ambient matter. The three solid lines with different colors indicate the critical accretion rates for three masses ($1.4\ M_\odot$,  $2.0\ M_\odot$, and $2.4\ M_\odot$, ) of the central object. The dashed lines show the estimated accretion rates onto a compact object with a mass of $1.4 \ M_\odot$ in CO cores of massive stars (see detail in the main text). Note that these accretion rates are proportional to the square of the mass of the compact object.}
 \label{fig:criticalmdot}
 \end{figure}

\section{Discussions}
\subsection{Implication for a new mechanism of supernovae}
As elaborated in section \ref{results}, an inspiral of a black-hole or a neutron star with a CO core may realize our calculation.  {The potential for these solutions to result in explosive burning is an interesting result of the calculations.} This phenomenon might be recognized as a super-Chandrasekhar Type-Ia supernova if the envelope outside the CO core had been stripped off before and both of the ejecta mass and the mass of radio active $^{56}$Ni exceed the Chandrasekhar limit. Observations for this class of supernovae indicate that the host galaxies are still undergoing star formation\citep{2006Natur.443..308H,2007ApJ...669L..17H,2010ApJ...713.1073S,2009ApJ...707L.118Y,2011MNRAS.412.2735T}, which is consistent with this speculation in which super-Chandrasekhar Type-Ia supernovae are of massive star origin. 

{Another possibility when explosive burning is realized as a result of the spiral-in of a neutron star is a partial ejection of the stellar material. This would happen when the explosive burning is not so strong.} Then the remaining stellar material would get higher temperatures and shallower densities. This may significantly reduce the accretion rate and result in a so called Thorne-Zytkow object \citep{1975ApJ...199L..19T}, of which the evolution is controlled by oxygen burning near the surface of a neutron star different from the original model controlled by hydrogen burning. Then the final outcome may be a violent ejection event associated with a catastrophic collapse of the accreting neutron star in a few years from the spiral-in. It is expected that the ejected material collides with the circumstellar matter formed by the detonation. This may be relevant to the recently discovered type Icn supernova\citep{galyam2021wcwo, perley2021type}.

{Thus it is critical to perform hydrodynamics simulations for this phenomenon to see whether a neutron star engulfed by a massive CO star results in explosive burning that significantly changes the structure of the CO star. }

\subsection{Implication for tidal disruption of CO white dwarf}
Based on the critical accretion rate obtained in the previous section, we are able to understand the reason why \cite{2019MNRAS.488..259F} did not see any significant influence from the nuclear burning to the accretion flow in tidal disruption of a white dwarf composed of C and O. The  accretion rate realized in the tidal disruption is of the order of 10$^{-3}\, M_\odot$ s$^{-1}$ at most. This is well below the value of the critical accretion for the carbon burning to significantly affect the accretion flow according to our results. Therefore their results are consistent with the notion of the critical accretion rate presented in this paper. 

\section{Conclusions}
We investigate the influence of nuclear burning on stationary spherically symmetric flows of matter composed of carbon and oxygen accreted by a neutron star or a black hole. We find that there are two types of such solutions depending on the accretion rate. One is a transonic flow that reaches the center and the other is a transonic flow that truncates at another sonic point. The critical accretion rates are also derived as functions of the mass of the central object and the specific enthalpy in the ambient matter. We compare this critical accretion rates with expected rates for a compact object inside some CO cores of massive stars, which are roughly estimated by a simple formula. Since the expected accretion rates are not too small compared with the critical accretion rate, we may expect that an engulfed neutron star (black hole) can generate a detonation wave that totally disrupts the CO star and result in a super-Chandrasekhar type Ia supernova or partially erupts the envelope and undergoes core collapse a few years later to emerge as a type Icn supernova. Of course, we need more detailed investigations such as multi-dimensional hydrodynamics simulations to obtain a firm conclusion.
\acknowledgments
TS would like to thank Kazumi Kashiyama and Daichi Tsuna on fruitful discussions on how to set up the problem and type Icn supernovae. This work is supported by JSPS KAKENHI Grant Numbers JP20H05639, JP22K03671, MEXT, Japan.

\bibliographystyle{aasjournal}
\bibliography{references}{}
\end{document}